\documentclass[12pt]{article}
\begin{document}

\title{THE SOLUTION OF THE MILNE PROBLEM FOR MAGNETIZED ATMOSPHERE}

\author{\it P.\,S.\,Shternin$^{1, 2}$, Yu.\,N.\,Gnedin$^1$ and N.\,A.\,Silant'ev$^3$}

\maketitle

\medskip

\begin{center}
$^1${\it Central Astronomical Observatory of Russian Academy of
sciences, St. Petersburg, Pulkovo, 196140, Russia}

$^2${\it St.Petersburg State Polytechnical University, Russia}

$^3${\it Instituto Nacional de Astrof\'{\i}isica, \'Optica y
Electr\'onica, Puebla, Apartado Postal 51 y 216, C. P. 72000,
M\'exico}\\
~

The paper is published in Astrophysics, {\bf 46}, 350 (2003).
\end{center}
~
~
\begin{abstract}
The numerical solution of the Milne problem for semi-infinite
plane-parallel magnetized electron atmosphere is obtained. It is
assumed that magnetic field is directed along the normal to the
atmosphere. The angular dependence, the polarization degree and
positional angle of outgoing radiation are presented in the tables
for various values of the Faraday rotation parameter and the
degree of absorption q=0, 0.2 and 0.4. We assume that magnetic
field $B\le 10^6 $ G when one can neglect the effects of circular
dichroism and take into account only the Faraday rotation effect.
\end{abstract}

\newpage

\section{Introduction}

When a magnetic field is present in hot electron atmospheres and
circumstellar shells, in accretion disks near quasars, and the nuclei of
active galaxies, radiation is subject to the Faraday rotation of its plane
of polarization. The angle of rotation $\psi $ is related to the parameters
of the medium and the path length $\it l$ over which the light travels by [1]:

\begin{equation}
\psi=\frac{1}{2}\delta \tau_T \cos\Theta,
\label{1}
\end{equation}

\noindent where $\tau_T=N_e\it l\sigma_T $ is the Thomson optical thickness of
the optical path, $\sigma_T=(8\pi/3)r_e^2\cong 6.65\cdot 10^{-25} cm^2$ is
the Thomson cross section, $r_e=e^2/m_e c^2\cong 2.82\cdot 10^{-13}cm$ is
the classical radius of electron, $N_e$ is the electron number density, and
$\Theta $ is the angle between the direction  $\bf n$ of the beam of light and
the magnetic field $\bf B$. The plane of polarization undergoes a right handed
rotation for $\Theta < 90^{\circ}$ and the opposite for $\Theta >90^{\circ}$ if
one views along the propagation direction of the light. The parameter $\delta $
is equal to the angle of rotation of the polarization plane on
a path $\tau_T=2$ along the magnetic field and is given by

\begin{equation}
\delta=\frac{3}{4\pi}\cdot \frac{\lambda}{r_e}\cdot
\frac{\omega_B}{\omega}\cong
0.8\lambda^2(\mu m)B(G).
\label{2}
\end{equation}

\noindent Here, $\lambda =2\pi c /\omega$ is the wavelength of the
radiation, $\omega =2\pi \nu$ is the angular frequency of the
light, $\omega_B =|e|B/m_e c$ is the electron cyclotron frequency
and $\omega_B/\omega\cong 0.93\cdot 10^{-8}\lambda(\mu m)B(G)$.

In the general case of elliptically polarized light, the Faraday rotation
describes the rotation of the polarization ellipse as the light passes through
a magnetized medium.

For atmospheres with $\omega_B/\omega \ll 1$, right and left circularly
polarized electromagnetic waves propagate in the medium independently at
phase velocities corresponding to the refractive indices $n_r$ and $n_{\it l}$.
A linearly polarized wave can be represented as a sum of right and left
polarized coherent waves and the difference in the phase velocities of these
waves leads to a rotation of the polarization plane
($\psi=0.5(\omega /c)l(n_l -n_r)$).  For $\omega_B/\omega \ll 1$ the scattering
cross sections for all the waves equal to the Thomson cross section $\sigma_T$.
We shall examine only this case here. For optical wavelengths
$(\lambda \approx 0.5\mu m)$, this means  $B\le 10^6 G$.

The Faraday rotation leads to depolarization of the radiation
since photons arriving from different optical depths have exposed
different rotations of their planes of polarization.

We are considering the so - called Milne problem, i.e., multiple
scattering of light in a semi - infinite plane - parallel
atmosphere where the sources of unpolarized radiation are located
at very high optical depth from the surface of the atmosphere.
Besides scattering of the light on electrons, we shall also
consider the intrinsic absorption of the light, the degree of
which we denote by $q (q=\sigma_a /(\sigma_a +\sigma_T)$, where
$\sigma_a$ is the intrinsic absorption cross section). The Milne
problem corresponds to the conditions in stellar atmospheres, as
well as the passage of light through optically thick circumstellar
shells and accretion disks.

Ambartsumyan's invariance principle [2] is usually used in solving
the Milne problem, that is, the angular distribution and
polarization of emerging radiation are independent of the adding
(or removal) of any layer to a semi - infinite atmosphere. Here
the intensity and polarization of the emerging radiation are
expressed in terms of so - called H-functions, which satisfy a
nonlinear integral equation in just angular variables.
Chandrasekhar [3] proposed an extremely efficient method for
solving this equation, the "fork" method, in which the successive
approximations represent the H-function with excess and
deficiency. For magnetized atmospheres the invariance principle
has been used [1,4,5] to obtain systems of nonlinear equations for
tensor H-functions. Unfortunately, there is no efficient method
for solving these extremely cumbersome systems of equations. For
the most interesting case of a conservative atmosphere ($q=0$),
the iteration procedure converges very poorly. Even in the
simplest case of a magnetic field perpendicular to the surface,
one must solve a system of six nonlinear equations.

For large values of the Faraday rotation parameter, $\delta \gg 1$, however,
the H-function technique can be used to obtain simple asymptotic solutions
of the Milne problem and of problems with power law and exponential
distributions of the sources in magnetized atmospheres [5]. These solutions
are suitable for arbitrary inclinations of the magnetic field to the outward
normal $\bf N$ to the atmosphere. Numerical solutions of these problems
have been obtained [7,8] for the case of zero magnetic field.

For $\delta \le 1$ and arbitrary inclination of magnetic field, all the
methods of solution lead to extremely cumbersome formulas and calculations.

For a magnetic field $\bf B$, directed along the normal $\bf N$ to the
surface of the atmosphere the calculations are much simpler, since the
problem has axial symmetry. In [9] the Milne problem for ${\bf B}\Vert \bf N$
has been solved by Monte Carlo method and in [10] by Feautrier method. However,
the results of these calculations, which were given in the form of graphs, are
sometimes rather different. In this paper the Milne problem for
${\bf B}\Vert \bf N$ is solved by the classical Chandrasekhar method [3] using
gaussian quadrature formulas to reduce the integro - differential transport
equation to a system of linear differential equations. By increasing the
order of the gaussian quadrature and comparing the results, one can estimate
the accuracy of the solutions that are obtained. We have attained accuracies
to the first four significant figures. This is the most accurate solution of
the Milne problem to date. In some cases our results differ by 10-15\% from
the less accurate results of [10]. It is extremely important that the
results obtained here can be used to estimate the accuracy of the simple
asymptotic formulas in [6].

\section{Solution of the Milne problem}

We shall consider the Milne problem for the case when the magnetic
field $\bf B$ is directed along the outward normal $\bf N$ to the
semi - infinite atmosphere. In the absence of a magnetic field the
problem reduces (see, [3]) to a system of coupled equations for
the intensity $I(\tau ,\mu)$ and the Stokes polarization parameter
$Q(\tau ,\mu )$ (either for the intensities of the radiation
polarized in the $(\bf nN)$ - plane or in the perpendicular plane,
$I_x(\tau ,\mu )$  and  $I_y(\tau ,\mu )$, respectively:
$I=I_x+I_y$, $Q=I_x-I_y $).  Here,  $\mu =\cos\vartheta $ is the
cosine of the angle between the propagation direction  $\bf n$ of
the light and the normal $\bf N$, $\tau $ is the optical
thickness, including absorption, taken from the surface into the
interior of the medium. As usual, we have chosen the $x$ axis of
the observer's system lying in the $({\bf nN})$ plane where $\bf
n$ is the direction to the telescope. Remind that the classical
Chandrasekhar solution for the emerging radiation ($\tau =0$)
gives an elongation of the angular distribution of
 $J(\mu =1)\equiv I(\mu =1)/I(\mu =0)=3.06$ and a peak value of the degree of
polarization for  $\mu =0$ of 11.71\%, while the oscillations of the electric
field vector of the radiation for all $\mu $ are perpendicular to
$({\bf nN})$-plane, i.e. $Q(\mu )<0$.

The presence of a magnetic field leads to the appearance of the
Stokes parameter $U(\tau ,\mu )=I_{x'}-I_{y'}$, where $x'$ and
$y'$ are the coordinate axes turned in the positive (right hand)
direction by $45^{\circ}$ from the basic $x$ and $y$ axes. The
parameter $U$ means that the plane of polarization is no longer
perpendicular to the plane  $({\bf nN})$. Remind that the angle of
inclination $\chi $ of this plane with respect to the
perpendicular to the plane $({\bf nN})$ is given by $\tan(2\chi
)=U/Q$. The azimuthal symmetry of this case, $\bf B\Vert N$, means
that $U$ does not contribute to the scattering on electrons, that
is, the integral term in the transport equation coincides with the
case $B=0$. In this situation  $U$ is completely determined by the
Faraday rotation process on leaving the atmosphere. This means
that the angle $\chi $ is taken in the right hand sense from the
plane of the oscillations without a magnetic field relative to the
line of sight at the telescope with a magnetic field directed
outward from the medium. For a magnetic field directed towards the
interior of the atmosphere, $\chi $ is taken in the opposite
direction.

The system of equations for $I(\tau ,\mu)$, $Q(\tau ,\mu)$ and $U(\tau ,\mu)$
according to the general formulas of [1,4] is

\[
\mu \frac{d}{d\tau}I(\tau ,\mu)=I(\tau ,\mu)-\frac{3}{16}(1-q)
\int\limits^1_{-1}d\mu' \left\{ [(3-\mu'^2)+
\mu^2(3\mu'^2-1)]I(\tau ,\mu')\right.+
\]
\begin{equation}
\left. (1-3\mu^2)(1-\mu'^2)Q(\tau ,\mu')\right\},
\label{3}
\end{equation}
\[
\mu\frac{d}{d\tau}Q(\tau ,\mu)=Q(\tau ,\mu)+(1-q)\delta \mu U(\tau ,\mu)-
\]
\begin{equation}
\frac{3}{16}(1-q)(1-\mu^2)\int\limits^1_{-1}d\mu'\left [(1-3\mu'^2)I(\tau ,\mu')
  +3(1-\mu'^2)Q(\tau ,\mu')\right ],
\label{4}
\end{equation}
\begin{equation}
\mu \frac{d}{d\tau }U(\tau ,\mu )=U(\tau ,\mu)-(1-q)\delta \mu Q(\tau ,\mu).
\label{5}
\end{equation}

\noindent Here $q$ is the degree of true absorption of the light,
$\tau $ is total optical thickness including absorption, and for
${\bf B\Vert N}$, $\cos\Theta =\mu $.

The boundary conditions for the system (3)-(5) are the usual ones:
$I(0,-\mu)=0$, $ Q(0,-\mu)=0$ and
$U(0,-\mu)=0$, that is, there is no radiation incident from outside.
In addition, it is assumed that none of the Stokes parameters have exponentially
increasing terms for  $\tau \to \infty $. Following Chandrasekhar's method [3], we
use gaussian quadratures to replace the integral terms with sums where the
parameters are taken at discrete points $\mu_i $: $I_i=I(\tau ,\mu_i)$,
$ Q_i=Q(\tau ,\mu_i)$ and $U_i=U(\tau ,\mu_i)$. The points
 $\mu_i$ are the roots of the Legendre polynomial $P_{2n}(\mu)$. The number $n$
determines the order of the gaussian quadrature formula. The system of
integro - differential equations (3)-(5) is thereby converted into a system of
ordinary differential equations:

\[
\mu_i \frac{d}{d\tau}I_i=I_i-\frac{3}{16}(1-q)\sum^{\pm n}_{j=\pm 1}a_j
\left \{[(3-\mu_j^2)+\mu_i^2(3\mu_j^2-1)]I_j+\right.
\]
\begin{equation}
\left. (1-3\mu_i^2)(1-\mu_j^2)Q_j\right\},
\label{6}
\end{equation}
\[
\mu_i\frac{d}{d\tau}Q_i=Q_i+(1-q)\delta\mu_i U_i-
\]
\begin{equation}
\frac{3}{16}(1-q)(1-\mu_i^2)\sum^{\pm n}_{j=\pm 1}a_j
\left [(1-3\mu^2_j)I_j+3(1-\mu_j^2)Q_j\right ],
\label{7}
\end{equation}
\begin{equation}
\mu_i\frac{d}{d\tau}U_i=U_I-(1-q)\delta\mu_i Q_i.
\label{8}
\end{equation}

\noindent Here, $\mu_i$ are the roots of the Legendre polynomial
($P_{2n}(\mu_i)=0$),$\mu_{-i}=-\mu_i$, and $a_i$ are the known weights of the
gaussian quadrature formula, with $a_{-i}=a_i$.

We seek a solution of the system (6)-(8) of the form

\begin{equation}
I_i=g_i\exp(-k\tau ),\quad Q_i=h_i\exp(-k\tau ),\quad U_i=f_i\exp(-k\tau ).
\label{9}
\end{equation}

\noindent Substituting (9) in (6)-(8) yields the formulas

\[
f_i=\frac{(1-q)\delta\mu_i}{1+k\mu_i}h_i,
\]
\[
g_i=\frac{\beta-\alpha\mu_i^2}{1+k\mu_i},
\]

\begin{equation}
h_i=\alpha\frac{(1-\mu_i^2)(1+k\mu_i)}{(1+k\mu_i)^2+[(1-q)\delta\mu_i]^2}.
\label{10}
\end{equation}

A homogeneous system of algebraic equations is derived from (6) and (7) for
finding the numbers $\alpha $ and $\beta $. The condition for this system to
be soluble, namely that the determinant equal to zero, yields the
characteristic equation for finding the eigenvalues $k$. We can only
determine the ratio $\alpha /\beta $ from the homogeneous second order system,
so that one of  $\alpha $ or $\beta $ remains unknown. This number is found
using the condition that the radiation flux leaving the atmosphere $F$ is
given.

As a result, we have obtained the angular distribution

\begin{equation}
J(\mu)=I(0,\mu)/I(0,0),
\label{11}
\end{equation}

\noindent the degree of polarization

\begin{equation}
p(\mu)=\frac{\sqrt{Q^{2}(0,\mu)+U^{2}(0,\mu)}}{I(0,\mu)},
\label{12}
\end{equation}

\noindent and the angle of inclination $\chi(\mu )$ of the electric field
oscillations of the radiation relative to a plane perpendicular to the plane
$({\bf nN})$,

\begin{equation}
\tan(2\chi )=\frac{U(0,\mu )}{Q(0,\mu )}.
\end{equation}

\noindent These quantities are listed in Tables 1--6 for a various
values of parameters $\delta $ and $q$.

\section{Conclusion}

We now analyze these results briefly and offer a qualitative explanation
of them. First of all, it is evident that the polarization of the radiation
$p(\mu )$ becomes ever more peaked as $\delta $ increases, with a maximum at
$\mu = 0$, i.e., in a direction perpendicular to the magnetic field. This is
a manifestation of the depolarization of the radiation owing to Faraday's
rotation, since the radiation emerging from the medium consists of the fluxes
of light undergoing different amounts of the Faraday rotation of their planes
of polarization. At the same time, as $\delta $ increases, there is an increase
in the angle $\chi $  by which the polarization plane of the emerging radiation
turns relative to the plane of polarization in the absence of a magnetic field,
i.e., relative perpendicular to the $({\bf nN})$ plane. In the limit
$\delta \to \infty $, the angle $\chi \to 45^{\circ }$. This behaviour of the
turning angle $\chi $ can be explained qualitatively as follows: the emerging
radiation mainly passes from the outer layer of the atmosphere with
 $\tau /\mu \approx 1$. According to (5), the Stokes parameter
$U(\mu )$ acquires the value -$\sim Q(\mu )(1-q)\delta \mu \tau /\mu $, which
leads to a ratio $U(\mu )/Q(\mu ) \sim (1-q)\delta \mu $. Thus, the turning
angle $\chi $ is given by

\begin{equation}
\tan(2\chi )\sim (1-q)\delta \mu . \label{14}
\end{equation}

For $(1-q)\delta \mu \gg 1$ the positional angle $\chi $ actually
tends to the limit  $45^{\circ }$. The Faraday rotation is
determined solely by the presence of free electrons along the path
of radiation, i.e., by the Thomson optical thickness
$\tau_T=(1-q)\tau $. For $q\to 1$, the outer layer of the
atmosphere with $\tau \approx 1$ contains too few electrons
($\tau_T \to 0 $) for the Faraday rotation to influence the
distribution of the polarization plane of emerging radiation. In
this case, we can neglect the parameter  $U(\tau ,\mu )$ and the
system (3)-(5) transforms into the ordinary equations of radiative
transfer without magnetic field. These qualitative arguments and
estimates are general in nature and not associated with the
specifics of the Milne problem. Thus, for arbitrary strongly
absorbing atmospheres ($q\to 1$) the Faraday rotation is
unimportant. The Milne problem is not interesting for highly
absorbing atmospheres, since the distribution of thermal radiation
sources, which is proportional to the distribution of absorbing
particles, plays a dominant role. It is known that radiation
emerging from highly absorbing layers is essentially unpolarized.

Polarization causes little change in the distribution of the emerging
radiation, even in the absence of a magnetic field. Thus, the Milne
problem with polarization taken into account (3) and (4) yields an elongation
$J(0)=3.06$, while solving (3) with the  $Q(\tau, \mu)$ term omitted (the
equation just for the intensity with the Rayleigh phase function) gives
$J(0)=3.02$. That is, the angular distributions are essentially the same
(see, Table 6).

The Faraday rotation causes depolarization of the radiation for all
the directions except perpendicular to the magnetic field. Thus, with
the increasing  $\delta $ the contribution of the polarization terms to the
formation of the angular distribution becomes ever smaller. Equation (3)
is transformed into a separate equation for just the intensity with the
Rayleigh phase function. Our tables show the gradual approach of the angular
distribution to this limiting form (see columns 3-5 of Table 6) as the
Faraday rotation parameter  $\delta $ increases.

The contribution of the polarization terms $Q$ and $U$ to the overall
polarization of the radiation emerging from the atmosphere is much greater
than their influence on the formation of the angular distribution. Thus,
the calculated degrees of polarization using known intensities of the radiation
are $9.37\%$ instead of $11.71\%$. This means that the difference
$11.71\%-9.37\%=2.34\%$ (or $20\%$ of the total polarization) is created by the
polarization terms. As can be seen from the Tables, the Faraday rotation
substantionally reduces the polarization and, for $\delta \gg 1$ the
contribution of the polarization terms to the degree of polarization itself
goes to zero. Here the intensity of radiation, itself, differs somewhat
from the case of zero magnetic field and is determined by the scalar transport
equation with the Rayleigh phase function (we are comparing the second and
third columns of Table 6). Thus, at the polarization maximum ($\mu =0$) we
obtain a slightly lower value of $9.14\%$ instead of $9.37\%$. Our values of
 $p(0)$ approach precisely this limit as $\delta \to \infty$ (see, Tables 1-3).

The simple asymptotic formulas in [6] for a number of standard
problems in radiation transport theory correspond to the
approximation in which the radiation intensity is determined by
the transport equation with the Rayleigh phase function, while the
polarization is treated as the result of single scattering of a
known radiation flux and its transformation by the Faraday
rotation. They give somewhat overestimated polarization values. A
comparison of the calculations using these formulas with the exact
solutions obtained here shows that for  $q=0$, with  $\delta = 10$
the asymptotic formulas yield polarization values with an relative
error of  $\approx 10\%$. For $\delta =5$ the error exceeds
$\approx 20\%$.

These simple formulas have the major advantage of providing an
analytical description of the polarization for an arbitrary
arrangement of the magnetic field in the atmosphere.

\section{Acknowledgments}

This work was supported by RFBR Grant No. 03-02-17223 and by Programs
of DFS and Presidium RAN and Federal Program of "Astronomy".

\newpage

\begin{table*}
\caption{The degree of polarization $p(\mu )$\%, the positional
angle of the polarization $\chi^{\circ }$, and angular
distribution  $J(\mu )$ of emerging radiation for $q=0$.}
\setlength{\tabcolsep}{0.12cm}
\begin{tabular}{llllllllllllllllllllll}
\noalign{\smallskip}
\hline
\noalign{\smallskip}
  & \multicolumn{3}{c}{$\delta =0$} & &
\multicolumn{3}{c}{1} &  & \multicolumn{3}{c}{2} &  &
\multicolumn{3}{c}{3} \\
\cline{2-4}\cline{6-8}\cline{10-12}\cline{14-16}
\noalign{\smallskip}
$\mu $~~~~~ & p & $\chi $ & J & & p & $\chi $ & J & & p & $\chi $ & J
& & p & $\chi $ & J \\
\noalign{\smallskip}
\hline
\noalign{\smallskip}
0      & 11.71  & 0~  & 1~  & & 11.56  & 0  & 1~  & & 11.26  & 0  & 1~  & &
10.98  & 0  & 1  \\
0.05   & 8.997  & 0~  & 1.1433~  & & 8.865  & 1.29  & 1.1437~  & &
8.497  & 2.537  & 1.1441~  & & 8.141  & 3.748  & 1.1442 \\
0.10   & 7.467  & 0~  & 1.2627~  & & 7.268  & 2.40  & 1.2635~  & &
6.842  & 4.650  & 1.2640~  & & 6.425  & 6.760  & 1.2642 \\
0.15   & 6.323  & 0~  & 1.3742~  & & 6.098  & 3.38  & 1.3752~  & &
5.633  & 6.439  & 1.3758~  & & 5.183  & 9.213  & 1.3760 \\
0.20   & 5.430  & 0~  & 1.4815~  & & 5.176  & 4.25 & 1.4827~  & &
4.693  & 7.962  & 1.4833~  & & 4.234  & 11.22  & 1.4832 \\
0.25   & 4.682  & 0~  & 1.5862~  & & 4.424  & 5.03  & 1.5875~  & &
3.939  & 9.267  & 1.5880~  & & 3.491  & 12.88  & 1.5877 \\
0.30   & 4.052  & 0~  & 1.6891~  & & 3.795  & 5.72  & 1.6905~  & &
3.322  & 10.39  & 1.6907~  & & 2.896  & 14.26  & 1.6902 \\
0.35   & 3.511  & 0~  & 1.7907~  & & 3.259  & 6.35  & 1.7920~  & &
2.809  & 11.37  & 1.7920~  & & 2.414  & 15.43  & 1.7912 \\
0.40   & 3.040  & 0~  & 1.8912~  & & 2.798  & 6.91  & 1.8926~  & &
2.377  & 12.23  & 1.8923~  & & 2.016  & 16.41  & 1.8911 \\
0.45   & 2.625  & 0~  & 1.9910~  & & 2.397  & 7.42  & 1.9923~  & &
2.008  & 12.98  & 1.9916~  & & 1.684  & 17.26  & 1.9901 \\
0.50   & 2.257  & 0~  & 2.0901~  & & 2.045  & 7.89  & 2.0913~  & &
1.692  & 13.65  & 2.0903~  & & 1.404  & 18.00  & 2.0885 \\
0.55   & 1.927  & 0~  & 2.1887~  & & 1.733  & 8.31  & 2.1898~  & &
1.417  & 14.24  & 2.1884~  & & 1.166  & 18.64  & 2.1862 \\
0.60   & 1.630  & 0~  & 2.2869~  & & 1.455  & 8.70  & 2.2879~  & &
1.178  & 14.78  & 2.2859~  & & 0.960  & 19.20  & 2.2834 \\
0.65   & 1.360  & 0~  & 2.3847~  & & 1.206  & 9.06  & 2.3855~  & &
0.967  & 15.26  & 2.3831~  & & 0.783  & 19.70  & 2.3802 \\
0.70   & 1.115  & 0~  & 2.4822~  & & 0.982  & 9.39  & 2.4829~  & &
0.780  & 15.69  & 2.4800~  & & 0.627  & 20.14  & 2.4766 \\
0.75   & 0.890  & 0~  & 2.5795~  & & 0.779  & 9.70  & 2.5799~  & &
0.613  & 16.08  & 2.5765~  & & 0.490  & 20.53  & 2.5728 \\
0.80   & 0.683  & 0~  & 2.6765~  & & 0.595  & 9.98  & 2.6767~  & &
0.465  & 16.44  & 2.6728~  & & 0.369  & 20.89  & 2.6687 \\
0.85   & 0.493  & 0~  & 2.7733~  & & 0.426  & 10.24  & 2.7733~  & &
0.331  & 16.77  & 2.7688~  & & 0.262  & 21.21  & 2.7643 \\
0.90   & 0.316  & 0~  & 2.8700~  & & 0.272  & 10.49  & 2.8697~  & &
0.210  & 17.07  & 2.8646~  & & 0.165  & 21.51  & 2.8597 \\
0.95   & 0.152  & 0~  & 2.9665~  & & 0.131  & 10.72  & 2.9659~  & &
0.100  & 17.35  & 2.9603~  & & 0.078  & 21.77  & 2.9550 \\
1      & 0      & 0~  & 3.0628~  & & 0      & 10.95      & 3.0620~  & &
0      & 17.60  & 3.0558~  & & 0      & 22.02  & 3.0500 \\
\noalign{\smallskip}
\hline
\end{tabular}
\end{table*}

\begin{table*}
\caption{The degree of polarization $p(\mu )$\%, the positional
angle of the polarization $\chi^{\circ }$, and the angular
distribution $J(\mu )$ of emerging radiation for $q=0$.}
\setlength{\tabcolsep}{0.12cm}
\begin{tabular}{llllllllllllllllllllll}
\noalign{\smallskip}
\hline
\noalign{\smallskip}
  & \multicolumn{3}{c}{$\delta =4$} & &
\multicolumn{3}{c}{5} &  & \multicolumn{3}{c}{6} &  &
\multicolumn{3}{c}{7} \\
\cline{2-4}\cline{6-8}\cline{10-12}\cline{14-16}
\noalign{\smallskip}
$\mu $~~~~~ & p & $\chi $ & J & & p & $\chi $ & J & & p & $\chi $ & J
& & p & $\chi $ & J \\
\noalign{\smallskip}
\hline
\noalign{\smallskip}
0      & 10.75  & 0  & 1~  & & 10.56  & 0  & 1~  & & 10.40  & 0  & 1~  & &
10.28  & 0  & 1  \\
0.05   & 7.802  & 4.913 & 1.1446~  & & 7.550  & 6.078  & 1.1447~  & &
7.336  & 7.218  & 1.1446~  & & 7.141  & 8.331  & 1.1445 \\
0.10   & 6.020  & 8.724  & 1.2646~  & & 5.712  & 10.64  & 1.2646~  & &
5.434  & 12.43  & 1.2644~  & & 5.175  & 14.11  & 1.2642 \\
0.15   & 4.751  & 11.71  & 1.3762~  & & 4.418  & 14.06  & 1.3761~  & &
4.120  & 16.17  & 1.3758~  & & 3.848  & 18.06  & 1.3755 \\
0.20   & 3.805  & 14.06  & 1.4834~  & & 3.477  & 16.64 & 1.4831~  & &
3.190  & 18.88  & 1.4826~  & & 2.936  & 20.82  & 1.4822 \\
0.25   & 3.083  & 15.94  & 1.5876~  & & 2.776  & 18.62  & 1.5872~  & &
2.515  & 20.89  & 1.5866~  & & 2.291  & 22.81  & 1.5860 \\
0.30   & 2.520  & 17.45  & 1.6899~  & & 2.243  & 20.17  & 1.6893~  & &
2.012  & 22.42  & 1.6885~  & & 1.818  & 24.29  & 1.6879 \\
0.35   & 2.074  & 18.69  & 1.7907~  & & 1.828  & 21.41  & 1.7899~  & &
1.627  & 23.62  & 1.7890~  & & 1.462  & 25.43  & 1.7882 \\
0.40   & 1.713  & 19.71  & 1.8903~  & & 1.499  & 22.41  & 1.8894~  & &
1.326  & 24.57  & 1.8883~  & & 1.186  & 26.33  & 1.8874 \\
0.45   & 1.418  & 20.57  & 1.9891~  & & 1.232  & 23.24  & 1.9880~  & &
1.085  & 25.35  & 1.9867~  & & 0.967  & 27.06  & 1.9857 \\
0.50   & 1.173  & 21.30  & 2.0871~  & & 1.014  & 23.94  & 2.0858~  & &
0.890  & 26.00  & 2.0844~  & & 0.790  & 27.66  & 2.0833 \\
0.55   & 0.968  & 21.93  & 2.1846~  & & 0.832  & 24.53  & 2.1831~  & &
0.728  & 26.55  & 2.1815~  & & 0.645  & 28.16  & 2.1803 \\
0.60   & 0.793  & 22.48  & 2.2815~  & & 0.679  & 25.03  & 2.2799~  & &
0.592  & 27.01  & 2.2781~  & & 0.524  & 28.59  & 2.2768 \\
0.65   & 0.643  & 22.95  & 2.3780~  & & 0.549  & 25.47  & 2.3762~  & &
0.478  & 27.41  & 2.3743~  & & 0.422  & 28.95  & 2.3728 \\
0.70   & 0.513  & 23.37  & 2.4742~  & & 0.437  & 25.85  & 2.4722~  & &
0.379  & 27.76  & 2.4701~  & & 0.334  & 29.27  & 2.4685 \\
0.75   & 0.399  & 23.74  & 2.5700~  & & 0.339  & 26.19  & 2.5679~  & &
0.294  & 28.06  & 2.5657~  & & 0.259  & 29.55  & 2.5639 \\
0.80   & 0.300  & 24.07  & 2.6656~  & & 0.254  & 26.49  & 2.6633~  & &
0.220  & 28.33  & 2.6609~  & & 0.194  & 29.79  & 2.6591 \\
0.85   & 0.212  & 24.37  & 2.7610~  & & 0.179  & 26.76  & 2.7585~  & &
0.155  & 28.57  & 2.7559~  & & 0.136  & 30.01  & 2.7540 \\
0.90   & 0.133  & 24.64  & 2.8561~  & & 0.113  & 27.00  & 2.8534~  & &
0.097  & 28.79  & 2.8507~  & & 0.085  & 30.20  & 2.8486 \\
0.95   & 0.063  & 24.88  & 2.9511~  & & 0.053  & 27.22  & 2.9482~  & &
0.046  & 28.98  & 2.9454~  & & 0.040  & 30.38  & 2.9432 \\
1      & 0      & 25.10  & 3.0459~  & & 0      & 27.41      & 3.0428~  & &
0      & 29.16  & 3.0398~  & & 0      & 30.53  & 3.0375 \\
\noalign{\smallskip}
\hline
\end{tabular}
\end{table*}

\begin{table*}
\caption{The degree of polarization $p(\mu )$\%, the positional
angle of the polarization $\chi^{\circ }$, and the angular
distribution
 $J(\mu )$ of emerging radiation for $q=0$.}
\setlength{\tabcolsep}{0.12cm}
\begin{tabular}{llllllllllllllllllllll}
\noalign{\smallskip}
\hline
\noalign{\smallskip}
  & \multicolumn{3}{c}{$\delta =8$} & &
\multicolumn{3}{c}{9} &  & \multicolumn{3}{c}{10} &  &
\multicolumn{3}{c}{100} \\
\cline{2-4}\cline{6-8}\cline{10-12}\cline{14-16}
\noalign{\smallskip}
$\mu $~~~~~ & p & $\chi $ & J & & p & $\chi $ & J & & p & $\chi $ & J
& & p & $\chi $ & J \\
\noalign{\smallskip}
\hline
\noalign{\smallskip}
0      & 10.17  & 0  & 1~  & & 10.08  & 0  & 1~  & & 10.01  & 0  & 1~  & &
9.173  & 0  & 1  \\
0.05   & 6.959  & 9.413  & 1.1444~  & & 6.787  & 10.46  & 1.1444~  & &
6.622  & 11.48  & 1.1444~  & & 1.540  & 37.72  & 1.1441 \\
0.10   & 4.931  & 15.67  & 1.2641~  & & 4.702  & 17.11  & 1.2640~  & &
4.487  & 18.46  & 1.2639~  & & 0.706  & 40.44  & 1.2631 \\
0.15   & 3.601  & 19.76  & 1.3753~  & & 3.377  & 21.28  & 1.3751~  & &
3.174  & 22.63  & 1.3749~  & & 0.430  & 41.37  & 1.3734 \\
0.20   & 2.713  & 22.51  & 1.4819~  & & 2.516  & 23.99 & 1.4816~  & &
2.342  & 25.29  & 1.4814~  & & 0.294  & 41.83  & 1.4792 \\
0.25   & 2.097  & 24.45  & 1.5856~  & & 1.931  & 25.86  & 1.5853~  & &
1.786  & 27.08  & 1.5850~  & & 0.215  & 42.11  & 1.5821 \\
0.30   & 1.654  & 25.87  & 1.6874~  & & 1.515  & 27.21  & 1.6869~  & &
1.396  & 28.36  & 1.6866~  & & 0.164  & 42.29  & 1.6830 \\
0.35   & 1.324  & 26.95  & 1.7876~  & & 1.208  & 28.23  & 1.7871~  & &
1.109  & 29.32  & 1.7867~  & & 0.128  & 42.42  & 1.7824 \\
0.40   & 1.070  & 27.79  & 1.8867~  & & 0.974  & 29.02  & 1.8861~  & &
0.892  & 30.06  & 1.8856~  & & 0.102  & 42.52  & 1.8807 \\
0.45   & 0.870  & 28.47  & 1.9849~  & & 0.790  & 29.65  & 1.9843~  & &
0.723  & 30.65  & 1.9837~  & & 0.082  & 42.60  & 1.9781 \\
0.50   & 0.710  & 29.02  & 2.0824~  & & 0.643  & 30.16  & 2.0817~  & &
0.588  & 31.13  & 2.0810~  & & 0.066  & 42.66  & 2.0747 \\
0.55   & 0.578  & 29.49  & 2.1793~  & & 0.523  & 30.59  & 2.1785~  & &
0.478  & 31.53  & 2.1778~  & & 0.053  & 42.71  & 2.1708 \\
0.60   & 0.469  & 29.88  & 2.2757~  & & 0.424  & 30.95  & 2.2748~  & &
0.387  & 31.86  & 2.2740~  & & 0.043  & 42.75  & 2.2664 \\
0.65   & 0.377  & 30.21  & 2.3716~  & & 0.341  & 31.26  & 2.3707~  & &
0.310  & 32.14  & 2.3699~  & & 0.034  & 42.78  & 2.3615 \\
0.70   & 0.299  & 30.50  & 2.4673~  & & 0.270  & 31.52  & 2.4662~  & &
0.246  & 32.39  & 2.4653~  & & 0.027  & 42.81  & 2.4564 \\
0.75   & 0.231  & 30.75  & 2.5625~  & & 0.209  & 31.76  & 2.5614~  & &
0.190  & 32.60  & 2.5605~  & & 0.021  & 42.83  & 2.5509 \\
0.80   & 0.173  & 30.98  & 2.6576~  & & 0.156  & 31.96  & 2.6564~  & &
0.142  & 32.79  & 2.6554~  & & 0.015  & 42.86  & 2.6451 \\
0.85   & 0.121  & 31.17  & 2.7524~  & & 0.109  & 32.14  & 2.7511~  & &
0.099  & 32.95  & 2.7500~  & & 0.011  & 42.88  & 2.7391 \\
0.90   & 0.076  & 31.35  & 2.8470~  & & 0.068  & 32.30  & 2.8456~  & &
0.062  & 33.10  & 2.8445~  & & 0.007  & 42.89  & 2.8329 \\
0.95   & 0.036  & 31.51  & 2.9414~  & & 0.032  & 32.44  & 2.9399~  & &
0.029  & 33.23  & 2.9387~  & & 0.003  & 42.91  & 2.9265 \\
1      & 0      & 31.65  & 3.0356~  & & 0      & 32.57  & 3.0341~  & &
0      & 33.35  & 3.0328~  & & 0      & 42.92  & 3.0200 \\
\noalign{\smallskip}
\hline
\end{tabular}
\end{table*}

\begin{table*}
\caption{The degree of polarization $p(\mu )$\%, the positional
angle of the polarization $\chi^{\circ }$, and the angular
distribution $J(\mu )$ of emerging radiation for $q=0.2$.}
\setlength{\tabcolsep}{0.12cm}
\begin{tabular}{llllllllllllllllllllll}
\noalign{\smallskip}
\hline
\noalign{\smallskip}
  & \multicolumn{3}{c}{$\delta =1$} & &
\multicolumn{3}{c}{5} &  & \multicolumn{3}{c}{10} &  &
\multicolumn{3}{c}{50} \\
\cline{2-4}\cline{6-8}\cline{10-12}\cline{14-16}
\noalign{\smallskip}
$\mu $~~~~~ & p & $\chi $ & J & & p & $\chi $ & J & & p & $\chi $ & J
& & p & $\chi $ & J \\
\noalign{\smallskip}
\hline
\noalign{\smallskip}
0      & 25.05  & 0  & 1~  & & 20.12  & 0  & 1~  & & 18.96  & 0  & 1~  & &
17.65  & 0  & 1  \\
0.05   & 22.95  & 1.188  & 1.1240~  & & 17.73  & 5.774  & 1.1248~  & &
15.67  & 11.01  & 1.1251~  & & 7.022  & 31.24  & 1.1270 \\
0.10   & 21.53  & 2.475  & 1.2337~  & & 15.44  & 11.42  & 1.2348~  & &
11.95  & 19.69  & 1.2357~  & & 3.491  & 37.32  & 1.2386 \\
0.15   & 20.28  & 3.868  & 1.3424~  & & 13.09  & 16.56  & 1.3436~  & &
8.877  & 25.61  & 1.3450~  & & 2.171  & 39.53  & 1.3480 \\
0.20   & 19.07 & 5.368  & 1.4542~  & & 10.85  & 20.96 & 1.4551~  & &
6.650  & 29.56  & 1.4567~  & & 1.499  & 40.67  & 1.4594 \\
0.25   & 17.85  & 6.973  & 1.5715~  & & 8.866  & 24.61  & 1.5716~  & &
5.068  & 32.27  & 1.5732~  & & 1.095  & 41.35  & 1.5752 \\
0.30   & 16.59  & 8.681  & 1.6963~  & & 7.190  & 27.57  & 1.6949~  & &
3.922  & 34.20  & 1.6962~  & & 0.828  & 41.81  & 1.6975 \\
0.35   & 15.27  & 10.49  & 1.8305~  & & 5.805  & 29.97  & 1.8269~  & &
3.070  & 35.64  & 1.8277~  & & 0.639  & 42.14  & 1.8279 \\
0.40   & 13.91  & 12.38  & 1.9761~  & & 4.671  & 31.92  & 1.9696~  & &
2.420  & 36.75  & 1.9694~  & & 0.499  & 42.38  & 1.9784 \\
0.45   & 12.50  & 14.35  & 2.1355~  & & 3.745  & 33.53  & 2.1249~  & &
1.913  & 37.62  & 2.1236~  & & 0.393  & 42.58  & 2.1211 \\
0.50   & 11.06  & 16.38  & 2.3112~  & & 2.986  & 34.86  & 2.2955~  & &
1.512  & 38.33  & 2.2928~  & & 0.309  & 42.73  & 2.2885 \\
0.55   & 9.605  & 18.47  & 2.5065~  & & 2.363  & 35.98  & 2.4843~  & &
1.189  & 38.91  & 2.4798~  & & 0.243  & 42.86  & 2.4734 \\
0.60   & 8.171  & 20.58  & 2.7253~  & & 1.850  & 36.93  & 2.6951~  & &
0.927  & 39.40  & 2.6883~  & & 0.189  & 42.96  & 2.6795 \\
0.65   & 6.779  & 22.69  & 2.9723~  & & 1.426  & 37.75  & 2.9323~  & &
0.713  & 39.81  & 2.9229~  & & 0.145  & 43.05  & 2.9111 \\
0.70   & 5.456  & 24.80  & 3.2537~  & & 1.077  & 38.45  & 3.2017~  & &
0.538  & 40.16  & 3.1891~  & & 0.109  & 43.13  & 3.1738 \\
0.75   & 4.226  & 26.87  & 3.5773~  & & 0.790  & 39.06  & 3.5108~  & &
0.394  & 40.47  & 3.4943~  & & 0.080  & 43.19  & 3.4750 \\
0.80   & 3.108  & 28.89  & 3.9534~  & & 0.555  & 39.60  & 3.8694~  & &
0.277  & 40.74  & 3.8483~  & & 0.056  & 43.25  & 3.8240 \\
0.85   & 2.120  & 30.84  & 4.3958~  & & 0.364  & 40.07  & 4.2907~  & &
0.182  & 40.98  & 4.2640~  & & 0.037  & 43.30  & 4.2339 \\
0.90   & 1.270  & 32.72  & 4.9235~  & & 0.211  & 40.50  & 4.7928~  & &
0.105  & 41.19  & 4.7594~  & & 0.021  & 43.35  & 4.7224 \\
0.95   & 0.563  & 34.51  & 5.5636~  & & 0.091  & 40.88  & 5.4018~  & &
0.046  & 41.38  & 5.3603~  & & 0.009  & 43.39  & 5.3147 \\
1      & 0      & 36.21  & 6.3553~  & & 0      & 41.22      & 6.1556~  & &
0      & 41.55  & 6.1043~  & & 0      & 43.42  & 6.0485 \\
\noalign{\smallskip}
\hline
\end{tabular}
\end{table*}

\begin{table*}
\caption{The degree of polarization $p(\mu )$\%, the positional
angle of the polarization $\chi^{\circ }$, and the angular
distribution $J(\mu )$ of emerging radiation for $q=0.4$.}
\setlength{\tabcolsep}{0.12cm}
\begin{tabular}{llllllllllllllllllllll}
\noalign{\smallskip}
\hline
\noalign{\smallskip}
  & \multicolumn{3}{c}{$\delta =1$} & &
\multicolumn{3}{c}{5} &  & \multicolumn{3}{c}{10} &  &
\multicolumn{3}{c}{50} \\
\cline{2-4}\cline{6-8}\cline{10-12}\cline{14-16}
\noalign{\smallskip}
$\mu $~~~~~ & p & $\chi $ & J & & p & $\chi $ & J & & p & $\chi $ & J
& & p & $\chi $ & J \\
\noalign{\smallskip}
\hline
\noalign{\smallskip}
0      & 39.89  & 0  & 1~  & & 33.20  & 0  & 1~  & & 31.45  & 0  & 1~  & &
29.49  & 0  & 1  \\
0.05   & 38.30  & 0.911  & 1.1055~  & & 31.28  & 4.499  & 1.1062~  & &
28.50  & 8.75  & 1.1066~  & & 15.00  & 28.51  & 1.1087 \\
0.10   & 37.00  & 1.924  & 1.2046~  & & 28.93  & 9.228  & 1.2054~  & &
23.90  & 16.74  & 1.2063~  & & 7.819  & 36.17  & 1.2105 \\
0.15   & 35.64  & 3.043  & 1.3081~  & & 25.98  & 13.92  & 1.3085~  & &
19.02  & 23.09  & 1.3100~  & & 4.902  & 39.11  & 1.3150 \\
0.20   & 34.13 & 4.276  & 1.4198~  & & 22.65 & 18.33  & 1.4195~  & &
14.82  & 27.81  & 1.4213~  & & 3.379  & 40.63  & 1.4265 \\
0.25   & 32.44  & 5.633  & 1.5431~  & & 19.24  & 22.32  & 1.5413~  & &
11.49  & 31.29  & 1.5432~  & & 2.454  & 41.55  & 1.5479 \\
0.30   & 30.55  & 7.124  & 1.6810~  & & 16.00  & 25.81  & 1.6769~  & &
8.933  & 33.88  & 1.6785~  & & 1.838  & 42.17  & 1.6824 \\
0.35   & 28.46  & 8.762  & 1.8370~  & & 13.09  & 28.81  & 1.8297~  & &
6.958  & 35.85  & 1.8306~  & & 1.401  & 42.61  & 1.8332 \\
0.40   & 26.16  & 10.56  & 2.0157~  & & 10.56  & 31.36  & 2.0038~  & &
5.421  & 37.39  & 2.0036~  & & 1.077  & 42.95  & 2.0045 \\
0.45   & 23.68  & 12.52  & 2.2223~  & & 8.407  & 33.53  & 2.2046~  & &
4.213  & 38.62  & 2.2027~  & & 0.830  & 43.21  & 2.2013 \\
0.50   & 21.04  & 14.65  & 2.4643~  & & 6.604  & 35.38  & 2.4387~  & &
3.254  & 39.61  & 2.4346~  & & 0.638  & 43.41  & 2.4302 \\
0.55   & 18.27  & 16.94  & 2.7509~  & & 5.110  & 36.95  & 2.7155~  & &
2.489  & 40.44  & 2.7084~  & & 0.486  & 43.58  & 2.7002 \\
0.60   & 15.45  & 19.40  & 3.0956~  & & 3.883  & 38.31  & 3.0474~  & &
1.876  & 41.13  & 3.0364~  & & 0.366  & 43.72  & 3.0236 \\
0.65   & 12.63  & 22.00  & 3.5168~  & & 2.883  & 39.48  & 3.4525~  & &
1.386  & 41.72  & 3.4365~  & & 0.270  & 43.84  & 3.4177 \\
0.70   & 9.908  & 24.71  & 4.0422~  & & 2.078  & 40.50  & 3.9573~  & &
0.995  & 42.23  & 3.9349~  & & 0.194  & 43.95  & 3.9085 \\
0.75   & 7.377  & 27.49  & 4.7137~  & & 1.438  & 41.39  & 4.6022~  & &
0.687  & 42.67  & 4.5718~  & & 0.134  & 44.04  & 4.5359 \\
0.80   & 5.123  & 30.29  & 5.5996~  & & 0.940  & 42.18  & 5.4541~  & &
0.449  & 43.05  & 5.4132~  & & 0.088  & 44.12  & 5.3652 \\
0.85   & 3.221  & 33.06  & 6.8179~  & & 0.562  & 42.88  & 6.6284~  & &
0.268  & 43.39  & 6.5742~  & & 0.052  & 44.18  & 6.5107 \\
0.90   & 1.723  & 35.75  & 8.5928~  & & 0.289  & 43.50  & 8.3468~  & &
0.138  & 43.69  & 8.2758~  & & 0.027  & 44.25  & 8.1930 \\
0.95   & 0.650  & 38.33  & 11.407~  & & 0.106  & 44.06  & 11.094~  & &
0.050  & 43.97  & 11.004~  & & 0.010  & 44.30  & 10.899 \\
1      & 0      & 40.76  & 16.531~  & & 0      & 44.56      & 16.170~  & &
0      & 44.21  & 16.075~  & & 0      & 44.35  & 15.962 \\
\noalign{\smallskip}
\hline
\end{tabular}
\end{table*}

\begin{table*}
\caption{Some solutions of the Milne problem without a magnetic field. The
first two columns give the Chandrasekhar values of the degree of polarization
$p(\mu )$\% and the angular distribution $J(\mu )$ of emerging radiation
for $q=0$. The third, fourth and fifth columns describe the angular 
distribution of the radiation obtained by solving the equation just for
the intensity with the Rayleigh phase function for $q=0,\, 0.2$, and
$0.4$, respectively. The last columns are our solutions of the Milne problem
for $q=0.2$, and $0.4$.}
\setlength{\tabcolsep}{0.12cm}
\begin{tabular}{llllllllllllllllllllll}
\noalign{\smallskip}
\hline
\noalign{\smallskip}
  & \multicolumn{2}{c}{$q=0 $} & &
\multicolumn{1}{c}{0} & & \multicolumn{1}{c}{0.2} & & 
\multicolumn{1}{c}{0.4} & & \multicolumn{2}{c}{0.2} & &
\multicolumn{2}{c}{0.4} \\
\cline{2-3}\cline{11-12}\cline{14-15}
\noalign{\smallskip}
$\mu $~~~~~ & p & J & & J & & J & & J & & p &  J  & & p & J \\  
\noalign{\smallskip}
\hline
\noalign{\smallskip}
0      & 11.71  & 1  & &~ 1  & & 1  & & 1  & &~ 28.63  & 1  & &~ 44.54 & 1  \\
0.05   & 8.979  & 1.1460  & &~ 1.1469  & & 1.1301  & & 1.1122  & &~ 26.49  &
1.1236  & &~ 42.93  & 1.1052  \\
0.10   & 7.448  & 1.2644  & &~ 1.2647  & & 1.2407  & & 1.2133  & &~ 25.04  &  
1.2333  & &~ 41.59  & 1.2045  \\
0.15   & 6.311  & 1.3755  & &~ 1.3746  & & 1.3496  & & 1.3173  & &~ 23.80  &
1.3424  & &~ 40.21  & 1.3085  \\ 
0.20   & 5.410  & 1.4826  & &~ 1.4801  & & 1.4606  & & 1.4284  & &~ 22.62  &
1.4549  & &~ 38.72  & 1.4212  \\
0.25   & 4.667  & 1.5871  & &~ 1.5828  & & 1.5761  & & 1.5495  & &~ 21.47  &
1.5732  & &~ 37.09  & 1.5459  \\
0.30   & 4.041  & 1.6898  & &~ 1.6835  & & 1.6981  & & 1.6836  & &~ 20.30  &
1.6994  & &~ 35.30  & 1.6858  \\
0.35   & 3.502  & 1.7913  & &~ 1.7829  & & 1.8282  & & 1.8340  & &~ 19.11  &
1.8355  & &~ 33.37  & 1.8445  \\
0.40   & 3.033  & 1.8918  & &~ 1.8810  & & 1.9684  & & 2.0047  & &~ 17.88  &
1.9836  & &~ 31.29  & 2.0265  \\
0.45   & 2.619  & 1.9915  & &~ 1.9783  & & 2.1208  & & 2.2009  & &~ 16.61  &
2.1459  & &~ 29.08  & 2.2375  \\
0.50   & 2.252  & 2.0906  & &~ 2.0773  & & 2.2878  & & 2.4290  & &~ 15.29  &
2.3254  & &~ 26.75  & 2.4849  \\
0.55   & 1.923  & 2.1892  & &~ 2.1709  & & 2.4723  & & 2.6981  & &~ 13.93  &
2.5252  & &~ 24.32  & 2.7783  \\
0.60   & 1.627  & 2.2873  & &~ 2.2665  & & 2.6778  & & 3.0202  & &~ 12.52  &
2.7493  & &~ 21.79  & 3.1315  \\
0.65   & 1.358  & 2.3851  & &~ 2.3616  & & 2.9088  & & 3.4128  & &~ 11.08  &
3.0029  & &~ 19.19  & 3.5635  \\
0.70   & 1.112  & 2.4826  & &~ 2.4564  & & 3.1709  & & 3.9017  & &~ 9.588  &
3.2920  & &~ 16.52  & 4.1023  \\
0.75   & 0.888  & 2.5798  & &~ 2.5508  & & 3.4712  & & 4.5267  & &~ 8.062  &
3.6249  & &~ 13.81  & 4.7909  \\
0.80   & 0.682  & 2.6768  & &~ 2.6450  & & 3.8193  & & 5.3529  & &~ 6.503  &
4.0122  & &~ 11.07  & 5.6988  \\
0.85   & 0.492  & 2.7736  & &~ 2.7389  & & 4.2281  & & 6.4945  & &~ 4.913  &
4.4680  & &~ 8.302  & 6.9457  \\
0.90   & 0.316  & 2.8703  & &~ 2.8327  & & 4.7151  & & 8.1718  & &~ 3.297  &
5.0119  & &~ 5.528  & 8.7577  \\
0.95   & 0.152  & 2.9667  & &~ 2.9263  & & 5.3059  & & 10.872  & &~ 1.658  &
5.6714  & &~ 2.758  & 11.619  \\
1      & 0      & 3.0631  & &~ 3.0197  & & 6.0376  & & 15.932  & &~ 0  &
6.4868  & &~ 0    & 16.786  \\
\noalign{\smallskip}
\hline
\end{tabular}
\end{table*}


\begin{thebibliography}{99}
\bibitem{1}A. Z. Dolginov, Yu. N. Gnedin, N. A. Silant'ev, {\it Propagation and polarization of radiation in cosmic media}, Gordon
\& Breach, New York (1995).
\bibitem{2}V. A. Ambartsumyan, Astron. Zh., {\bf 19}, 1 (1942).
\bibitem{3}S. Chandrasekhar, {\it Radiative transfer}, Clarendon Press, Oxford (1950).
\bibitem{4}N. A. Silant'ev, Astrophys. Space Sci., {\bf 82}, 363 (1982).
\bibitem{5}N. A. Silant'ev, J Quant. Spectr. Radiat. Transfer, {\bf 52}, 207 (1994).
\bibitem{6}N. A. Silant'ev, Astronomy \& Astrophysics, {\bf 383}, 326 (2002).
\bibitem{7}V. M. Loskutov, V. V. Sobolev, Astrophysics, {\bf 15}, 241 (1979).
\bibitem{8}N. A. Silant'ev, SvA, {\bf 24}, 341 (1980).
\bibitem{9}E. Agol, O. Blaes, MNRAS, {\bf 282}, 965 (1996).
\bibitem{10}E. Agol, O. Blaes, C. Ionescu-Zanetti, MNRAS, {\bf 293}, 1 (1998).
\end{thebibliography}
\end{document}